\documentclass[a4]{article}
\usepackage{authblk}
\usepackage{graphicx}
\usepackage{amsmath}
\usepackage{amsfonts}
\begin{document}
\title{Investigation of the interaction of some astrobiological molecules
with the surface of a graphite (0001) substrate. Application to the CO, HCN, 
H$_2$O and H$_2$CO molecules.}
\author{Azzedine Lakhlifi\footnote{
Institut Universitaire de Technologie, Belfort-Montb\'{e}liard. \\
E-mail : azzedine.lakhlifi@obs-besancon.fr\\
Institut Utinam - UMR CNRS 6213  Universit\'{e} de Franche-Comt\'{e} 
Observatoire de Besan{\c c}on \\ 41 bis avenue de l'Observatoire - BP 1615 -
25010 Besan{\c c}on Cedex, France}  \and John P. Killingbeck\footnote{
Mathematics Department, University of Hull, Hull HU6 7RX, UK}}
\maketitle
\begin{abstract}
Detailed semi-empirical interaction potential calculations are performed
to determine the potential energy surface experienced by the molecules CO, HCN, H$_2$O
and H$_2$CO, when adsorbed on the basal plane (0001) of graphite at low temperature.
The potential energy surface is used to find the equilibrium site and configuration
of a molecule on the surface and its corresponding adsorption energy. The diffusion
constant associated with molecular surface diffusion is calculated for each molecule.
\end{abstract}
\section{Introduction}
Observations in the cold regions of the interstellar medium (ISM) with typical temperatures 
of about 10 - 20 K have revealed the presence of clouds which contain high densities 
(more than 10$^3$ particle per cm$^3$) of atoms, radicals and simple molecules such as H$_2$, CO, 
CO$_2$, HCN, H$_2$O (the most abundant molecule), NH$_3$, H$_2$CO, CH$_4$, CH$_3$OH, 
as well as small solids typically composed of amorphous H$_2$O, silicates and carbon grains in the 
form of graphite, amorphous structures and polycyclic aromatic hydrocarbon (PAH) 
molecules \cite{btd1988,jsmgw1989}.

It is nowadays thought that the dominant mechanism for the formation of molecules 
is provided by surface-catalysed chemical reactions on low temperatures interstellar dust grains. 
For instance, Watanabe and Kouchi \cite{nwak2002} have experimentally produced the formaldehyde H$_2$CO and 
methanol CH$_3$OH molecules by successive hydrogenation of CO on the surface of icy grains at 10 K, 
as was theoretically proposed by Tielens et al. \cite{aggmtdcbw1997}. Recent laboratory experiments 
reported that more complex molecules such as glycine and other 
amino acids (which play an important role in the origin and evolution of life) as well as other 
organic species were formed from ultraviolet (UV) irradiation of the analogues of icy 
interstellar grains at low temperatures (\textit{T} $\sim$ 15 K) and containing small amounts 
of HCN, NH$_3$, H$_2$CO, and CH$_3$OH species \cite{mpbjpdsasgwclja2002,gmmcujmwasbbaashrwhptabjmg2002}.

Theoretical quantum-chemical calculations on the synthesis of amino acids were reported by 
Woon \cite{dew2002} for UV-irradiated astrophysical ices and by Mendoza et al. \cite{cmfrgmlsr2004} 
for ice analogues on the surface of a coronene (PAH) molecule.

Investigations of the chemical processes leading to the production of amino acids on 
the surface of grains require a knowledge of the interactions between the molecules 
participating in these processes on the surface, and also of their adsorption characteristics (such as the 
equilibrium configurations and diffusion constants) when they are isolated on the grain surface.

The interaction of a single water molecule with a graphite sheet and with PAH molecules like the 
fused-benzene molecules (fbz)$_n$ (C$_6$H$_6$, C$_{24}$H$_{12}$, C$_{54}$H$_{18}$, C$_{96}$H$_{24}$,... 
for \textit{n} = 1, 7, 19, 37,..., respectively) has been theoretically studied using \textit{ab-initio} 
methods to determine the binding energies and equilibrium configurations \cite{avwas1992,rlgrab1964,
df1999,dfkdj2000,cslrqzstlmetfljw2005,slvrcttapdcl2008,stkhtummkt2000,uzmppk2004}.
For H$_2$O adsorbed on a graphite sheet, the calculated binding energies by various authors are 
-187 \cite{rlgrab1964}, -72 \cite{avwas1992}, and -109 meV \cite{kkkdj2003}. The 
available experimental value is -156 meV \cite{nnaavk1970}.

For the water-benzene complex, many theoretical \textit{ab-initio} calculations were reported. The 
calculated binding energies are around -120 \cite{stkhtummkt2000,uzmppk2004,slvrcttapdcl2008}, 
-169 meV \cite{df1999}, and -99 meV \cite{cslrqzstlmetfljw2005}. The available experimental 
values are -98 \cite{bmcjrgeaw1995}, and -106 meV \cite{acmmidfpmpgvbpdppm1998}.

Feller and Jordan \cite{dfkdj2000} have reported for water-(fbz)$_n$ complex up to 
\textit{n} = 37 an energy binding -251 meV which is twice as large as the recently 
reported value -126 meV by Lin et al. \cite{cslrqzstlmetfljw2005}.

In the present work we study the adsorption properties of several of the molecules which are 
associated with the synthesis of amino acids. We use the graphite (0001) surface as a model
substrate to represent the surface of an interstellar dust grain and we consider the surface
adsorption of the various molecules on the model surface. We use an approach which
has already been developed to calculate the infrared spectra of the NH$_3$ ammonia molecule adsorbed 
on a variety of surfaces \cite{alspcgaa1995,alcg1996,al2000,aljpk2005}.
The basic procedure involves the calculation of the potential energy surface for the
adsorbed molecule as it moves on the adsorbing surface. The environmental effect on
the inversion spectrum of NH$_3$, for example, can be simulated by changes in the depth
or curvature of an effective double well potential. Depending on the substrate surface
it may happen that, for the NH$_3$ molecule, the central barrier height increases (giving
reduced splittings within each of the lowest two vibrational levels) or an asymmetry
appears between the two potential wells. Even a slight asymmetry can inhibit the
inversion motion in NH$_3$, so that the eigenstates are effectively localized on one side
or the other of the central potential barrier. These earlier calculations with NH$_3$ have
given us experience of the calculational techniques involved and they led to calculated
spectra which were in agreement with the experimental ones.

The isosteric heat of adsorption for NH$_3$ adsorbed on the graphite substrate calculated 
in our model was 178 meV \cite{aljpk2005}, in agreement with the experimental 
value of 172 meV given by Avgul and Kiselev \cite{nnaavk1970} at a temperature \textit{T} = 195 K.

The various parts of section 2 of this work describe the important contributions
to the potential energy of the adsorbed molecule on the adsorbing surface. Section 3
describes the minimization process needed to find the possible equilibrium positions of
the adsorbed molecule on the surface and gives details of the observables which are to be
calculated. Section 4 gives the results obtained for the several molecules studied; typical
potential surfaces are shown in the figures, to augment the sometimes complicated verbal
description of the equilibrium positions of the molecule.

Two tables set out some important calculated numerical parameters for the various
adsorbed molecules. An appendix sets out some of the mathematical transformations
needed to transform various quantities between the internal molecular axes and the
chosen reference set of axes needed to give a unified description of the various quantities.
\section{Theoretical background}
An adsorbed species can be either physically or chemically bonded to the surface of the
substrate crystal. In physisorption the interaction is due to the comparatively weak
van der Waals-London and electric field-multipole forces, and the interaction potential
energy is fairly low ($\leq$ 0.5 eV), with no charge exchange between the adsorbed species
and the substrate.
\subsection{The physisorbed molecule on a graphite substrate}
In this work we consider a single molecule such as CO, HCN, H$_2$O or H$_2$CO physisorbed
on a graphite basal plane (0001) at low temperatures.

The graphite crystal is extensively used as a model to simulate the behaviour
of interstellar grains in studies of the physical and chemical behaviour of adsorbed
molecular species. Experimentally detected absorption features due to carbon grains
have been attributed to the presence of small spherical graphite crystals of size 
$\sim$ 200 $\AA$ ̊in interstellar dust clouds \cite{btd1988,jsmgw1989}.

In order to reduce the required numerical calculation times, the graphite crystal is
modelled by using ten planes with radii $\sim$ 35 $\AA$ ($\sim$ 1500 carbon atom per plane)̊. 
This does not significantly influence the interaction potential energy between the molecule 
and the substrate (giving an error of less than 2 percent).

Figure 1 shows the geometrical characteristics of a molecule (H$_2$CO for instance)
adsorbed on the graphite surface.

To perform the calculations, we need to define:

\textit{i}) an absolute frame (\textbf{O,X,Y,Z}) connected to the surface of the substrate, with
respect to which all of the vectors and the tensors associated with the molecule and the
graphite carbon atoms must be described;

\textit{ii}) a frame (\textbf{G,x,y,z}) tied to the molecule with centre of mass G, and with respect
to which the molecular vibrational motions are still described. The rotational matrix
transformation from the absolute frame to the molecular frame is given in Appendix
A, in terms of the molecular orientational degrees of freedom $\Omega = (\varphi,\theta,\chi)$ (precession, nutation, and proper rotation Euler angles).
\begin{figure}
\includegraphics[width=14cm]{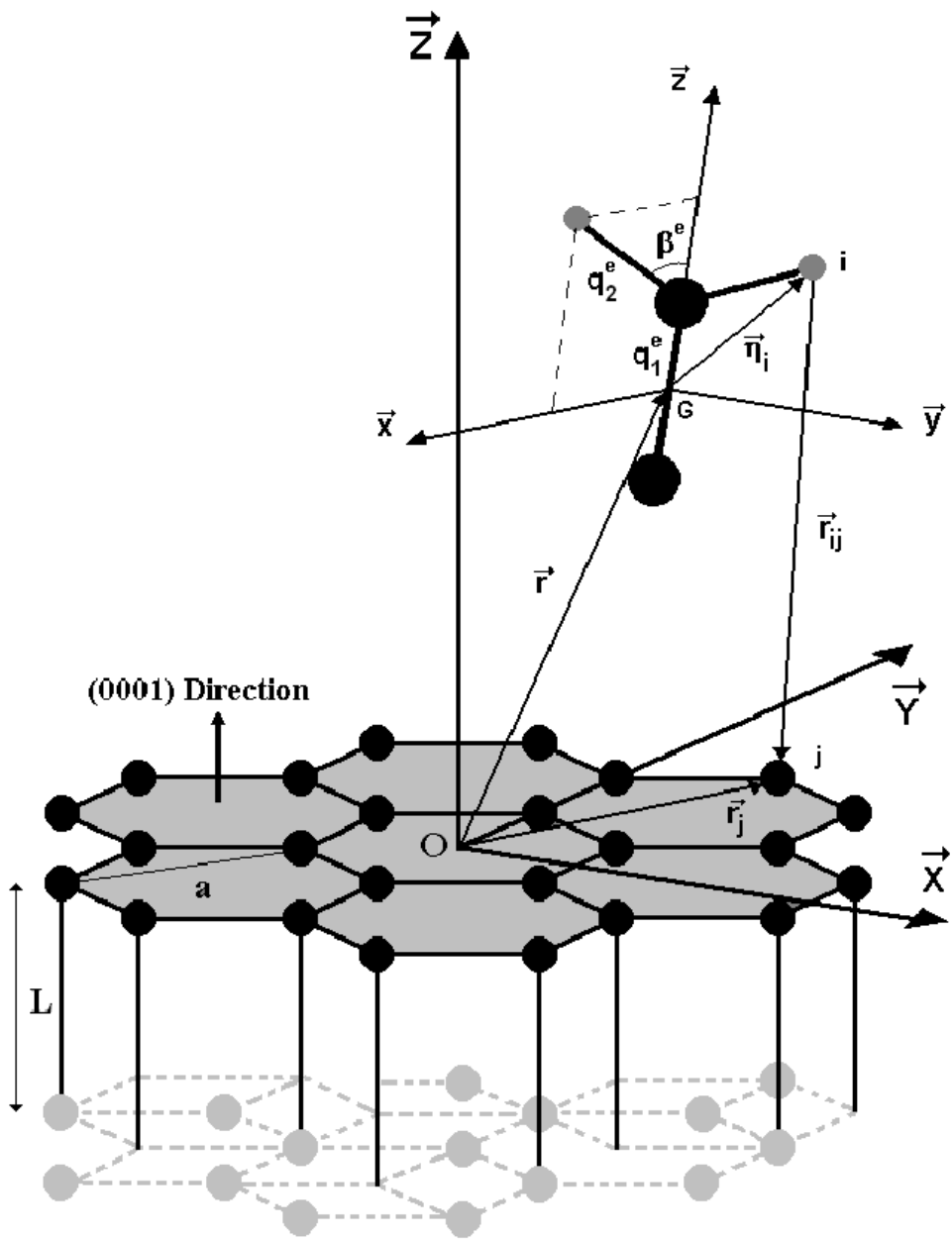}
\caption{Geometrical characteristics of an H$_2$CO molecule adsorbed on the graphite (0001) 
substrate. (\textbf{O,X,Y,Z}) and (\textbf{G,x,y,z}) represent the surface absolute frame 
and the molecular frame, respectively. \textit{L} and \textit{a} are the interlayer spacing 
and the distance between two next-nearest neighbour carbon atoms of the graphite and 
q$_{\text{1}}^{\text{e}}$, q$_{\text{2}}^{\text{e}}$ and $\beta^{\text{e}}$ define the 
equilibrium internal coordinates of the adsorbed molecule.}
\end{figure}
\subsection{Interaction potential energy}
The interaction potential energy $V_{\text{MG}}$ between an adsorbed molecule and the graphite
substrate can be written as the sum of three parts
\begin{equation}
V_{\text{MG}} = V_{\text{LJ}}+V_{\text{E}}+V_{\text{I}}, \label{eq1}
\end{equation}
\noindent which are detailed below.
\subsubsection{Quantum contributions}
The first term $V_{\text{LJ}}$ in Eq. (\ref{eq1}) is generally treated by
using semi-empirical 12-6 Lennard-Jones (LJ) pairwise atom-atom potentials to describe
the short-range atomic repulsion energy arising from the overlap of the electronic clouds
of the adsorbate and the substrate at very short distances and also the long-range many-
body dispersion energy arising from the lowering of the quantum zero-point energy of
the system by the correlated fluctuations of the adsorbate and substrate atomic dipoles.

However, it is widely known that carbon atoms in the graphite basal plane (0001)
exhibit anisotropic polarizability effects which modify the spherical shape of these atoms.
To account for these anisotropies, Carlos and Cole \cite{wecmwc1982} proposed a modified expression
for $V _{\text{LJ}}$ , of the form
\begin{eqnarray}
V_{\text{LJ}}=\sum\limits_j\sum\limits_i4\epsilon _{ij}
\left\{\left( \frac{\sigma _{ij}}{\left| \textbf{r}_{ij}\right| }\right) ^{\text{12}}
\left[ 1+\gamma _R(1-\frac 65\cos ^2\theta _{ij})\right] - \right. \nonumber \\ 
\left. \left( \frac{\sigma _{ij}}{\left| \textbf{r}_{ij}\right| }\right) ^{\text{6}}
\left[ 1+\gamma _A(1-\frac 32\cos ^2\theta _{ij})\right]\right\} .  \label{eq2}
\end{eqnarray}

In this equation, \textbf{r}$_{ij}$ is the distance vector between the \textit{i}th atom 
of the molecule and the \textit{j}th carbon atom of the graphite substrate (Figure 1), 
while $\epsilon _{ij}$ and $\sigma _{ij}$ are the mixed LJ potential parameters, obtained 
from the usual Lorentz-Berthelot combination rules 
$\epsilon _{ij} = \sqrt{\epsilon _{ii}\epsilon _{jj}}$ and 
2$\sigma _{ij} = \sigma _{ii}+\sigma _{jj}$.
The two coefficients $\gamma _{R}$ and $\gamma _{A}$ specify the modifications to 
the repulsion and dispersion parts of the energy, respectively, and $\theta _{ij}$ 
is the angle between the surface normal axis and the \textbf{r}$_{ij}$ vector. 
The various molecule-graphite mixed parameters are given in Table 1. Note that the LJ parameters 
used in this work have been improved several times in our other works.

\begin{table} 
\begin{center}
\caption{Lennard-Jones parameters for atom-graphite combinations.} 
\begin{tabular}{||c|c|c|c|c||}
\hline \hline
Gr - Atom & $\epsilon $ (meV) & $\sigma $ (\AA ) & $\gamma _{\text{A}}^{(a)}$
& $\gamma _{\text{R}}^{(a)}$ \\ 
\hline
Gr - H & 2.265 & 2.965 & 0.4 & -0.54 \\ 
Gr - O & 3.450 & 3.141 & 0.4 & -1.05 \\ 
Gr - C & 2.981 & 3.305 & 0.4 & -1.05 \\ 
Gr - N & 2.811 & 3.390 & 0.4 & -1.05 \\
\hline 
\end{tabular} 
\end{center}
\raggedright $(a)$ From Refs. \cite{wecmwc1982,gvmwc1984,fyhlwbser1992,fyhlwb1995,siimelv1992}.
\end{table}

\subsubsection{Electrostatic and induction contributions}
It is well known that the bonds in the graphite crystal give rise to aspherical atomic 
charge distributions which produce multipole electrostatic fields with rapid spatial 
variations external to the graphite \cite{avwas1992}. This phenomenon is modelled by introducing axially 
symmetric quadrupole moment tensors $\Theta ^{\text{c}}$ at each carbon atom site.

The quadrupole moment tensor $\Theta ^{\text{c}}$ of the \textit{j}th carbon atom 
produces the electrical potential
\begin{equation}
\mathbf{\Phi}^j(\mathbf{r})=\frac 13L_{\text{p}}\times \mathbf{\nabla} \mathbf{\nabla} \left(\frac
1{\left| \textbf{r}_j-\mathbf{r}\right| }\right) :\mathbf{\Theta} ^{\text{c}},  \label{eq3}
\end{equation}

\noindent on the molecule at its centre of mass position \textbf{r}. In turn, the multipole moments 
($\mu$, $\Theta$,...) of the molecule produce the electrical potential
\begin{equation}
\mathbf{\Phi} ^M(\mathbf{r}_j)=L_{\text{p}}\times \mathbf{\nabla} \left( \frac 1{\left| \mathbf{r}%
_j-\mathbf{r}\right| }\right) .\mathbf{\mu} +\frac 13L_{\text{p}}\times \mathbf{\nabla} \mathbf{\nabla}
\left( \frac 1{\left| \mathbf{r}_j-\mathbf{r}\right| }\right) :\mathbf{\Theta} + ...,  \label{eq4}
\end{equation}
\noindent on the \textit{j}th carbon atom at position \textbf{r}$_j$.

In Eqs. (\ref{eq3}) and (\ref{eq4}), $L_{\text{p}}$ is a screening factor, equal to 1 for the 
surface plane and to 2/$\left( \varepsilon +1\right) $ for the internal planes (p $\geq$ 2), where $\varepsilon$ is the static dielectric constant of the graphite crystal, which we take to be 2.8.

The electrostatic and induction contributions to the interaction potential energy $V_{\text{MG}}$
are
\begin{equation}
V_{\text{E}}=\sum\limits_j\mu .\nabla \Phi ^j(\text{r})+\frac
13\sum\limits_j\Theta :\nabla \nabla \Phi ^j(\text{r})+...  \label{eq5}
\end{equation}
\begin{equation}
V_{\text{I}}=-\frac 12\sum\limits_j\left[ \nabla \Phi ^M(\text{r}_j):\alpha
^{\text{c}}:\nabla \Phi ^M(\text{r}_j)\right] -\frac
12\sum\limits_{j,j^{\prime }}\left[ \nabla \Phi ^j(\text{r}):\alpha :\nabla
\Phi ^{j^{\prime }}(\text{r})\right] , \label{eq6}
\end{equation}
\noindent where $\alpha$ and $\alpha ^{\text{c}}$ are the polarizability tensors of the molecule and of the graphite carbon atoms, respectively.

The elements of the polarizability $\alpha^{\text{c}}$ and quadrupole moment 
$\Theta^{\text{c}}$ tensors (Eqs. (\ref{eq3}-\ref{eq6})) of the carbon atoms are given 
in the absolute frame (\textbf{O,X,Y,Z}) by
\begin{equation}
\alpha _{\text{XX}}^{\text{c}}=\alpha _{\text{YY}}^{\text{c}}=\alpha _{\perp
}^{\text{c}} \ \ \ \  \text{  and   } \ \ \ \  \alpha _{\text{ZZ}}^{\text{c}}=\alpha _{\parallel
}^{\text{c}}, \label{eq7}
\end{equation}
\begin{equation}
\Theta _{\text{XX}}^{\text{c}}=\Theta _{\text{YY}}^{\text{c}}=-\frac
12\Theta _{\text{ZZ}}^{\text{c}}=-\frac 12\Theta ^{\text{c}}, \label{eq8}
\end{equation}
\noindent with all other elements being zero.

Moreover, it must be noticed that the molecular frame corresponds, for each of
the studied molecules, to its principal frame of inertia, in which only the z-component
of the dipole moment $\mu$ vector and the three diagonal elements of the polarizability
$\alpha$ and quadrupole moment $\Theta$ tensors are non-vanishing elements. In the potential energy calculations these quantities must be expressed in the absolute frame by using a
rotational matrix transformation (Appendix A).

The various geometrical and electrical parameters of the graphite substrate and of
the studied molecules are given in Table 2.

\begin{table} 
\begin{center}
\caption{Molecular and graphite parameters: internal bonds and angles, dipole moments, 
polarizabilities, quadrupole moments, and rotational constants for the molecules studied.} 
\begin{tabular}{||c|c|c|c|c|c|c||}
\hline \hline
Molecule & CO & HCN & H$_2$O & H$_2$CO & Graphite$^{(a)}$ &  \\ 
\hline
\{q$^{\text{e}}$\} (\AA ) & 1.144 & 1.150 & 0.958 & 1.198 & $L$ (\AA ) & 3.36 \\ 
& - & 1.079 & - & 1.121 & $a$ (\AA ) & 2.46 \\ 
$\beta ^{\text{e}}$ (deg) & - & - & 54.7 & 58.1 &  &  \\ 
$\mu $ (D) & 0.112 & 2.986 & 1.855 & 2.330 &  &  \\ 
$\alpha _{\textrm{xx}}$ (\AA $^3$) & 1.63 & 2.05 & 1.53 & 2.51 & $\alpha _{%
%%\textrm{\perp }}^{\text{c}}$ (\AA $^3$) & 1.44 \\ 
\perp }^{\text{c}}$ (\AA $^3$) & 1.44 \\ 
$\alpha _{\textrm{yy}}$ (\AA $^3$) & 1.63 & 2.05 & 1.42 & 1.94 & $\alpha _{%
%\text{\parallel }}^{\text{c}}$ (\AA $^3$) & 0.41 \\ 
\parallel }^{\text{c}}$ (\AA $^3$) & 0.41 \\ 
$\alpha _{\textrm{zz}}$ (\AA $^3$) & 2.60 & 3.66 & 1.47 & 2.90 &  &  \\ 
$\Theta _{\textrm{xx}}$ (D\AA ) & 1.02 & -2.20 & 2.63 & 1.91 & $\Theta ^{\textrm{%
c}}$ (D\AA ) & 1.0 \\ 
$\Theta _{\textrm{yy}}$ (D\AA ) & 1.02 & -2.20 & -2.50 & -5.01 &  &  \\ 
$\Theta _{\textrm{zz}}$ (D\AA ) & -2.05 & 4.40 & -0.13 & 3.09 &  &  \\ 
A (cm$^{-1}$) & 1.93 & 1.48 & 27.33 & 1.22 &  &  \\ 
B (cm$^{-1}$) & 1.93 & 1.48 & 14.57 & 1.13 &  &  \\ 
C (cm$^{-1}$) & - & - & 9.49 & 9.40 &  &  \\
\hline
\end{tabular} 
\end{center}
\raggedright $(a)$ From Refs. \cite{wecmwc1982,gvmwc1984,fyhlwbser1992,fyhlwb1995,siimelv1992}.
\end{table}

\subsubsection{Separation of $V_{\text{MG}}$}
The distance vector \textbf{r}$_{ij}$ in Eq. (\ref{eq2}) can be expressed as
\begin{equation}
\mathbf{r}_{ij}=\mathbf{r}_{j}-\mathbf{r}-\mathbf{\eta}_{i}, \label{eq9}
\end{equation}
\noindent where \bf{r} and \bf{r}$_{j}$ are the position vectors of the molecular 
centre of mass and the \textit{j}th graphite carbon atom, respectively.

The vectors {$\eta_{i}$} characterize the positions of the atoms in the molecule. They, as
well as the molecular electrical parameters $\mu$, $\Theta$, $\alpha$, depend on the molecular internal
vibrational motions. They could be written, with respect to the (\textbf{G,x,y,z}) frame, as
\begin{eqnarray}
\eta _i &=&\eta _i^{\text{e}}+\sum\limits_\nu \textbf{a}_i^\nu Q_\nu +..., \nonumber \\
\mu  &=&\mu ^{\text{e}}+\sum\limits_\nu \textbf{b}^\nu Q_\nu +..., \nonumber \\
\Theta  &=&\Theta ^{\text{e}}+\sum\limits_\nu \textbf{c}^\nu Q_\nu +..., \nonumber \\
\alpha  &=&\alpha ^{\text{e}}+\sum\limits_\nu \textbf{d}^\nu Q_\nu +.... \label{eq10}
\end{eqnarray}
In these expressions the superscript e refers to the equilibrium internal configuration
of the molecule (rigid molecule) and $\textbf{a}_i^\nu$, $\textbf{b}^\nu$, $\textbf{c}^\nu$, and 
$\textbf{d}^\nu$ are the first derivatives of $\eta_{i}$, $\mu$, $\Theta$, and $\alpha$ with 
respect to the normal coordinate $Q_\nu$ associated with the $\nu$th molecular vibrational 
mode with frequency $\omega_\nu$.

However, the present work is devoted to finding the adsorption energies and
diffusion constants of single molecules adsorbed on the graphite substrate. Therefore, we
can assume: \textit{i}) the substrate to be rigid ($\{\textbf{r}_{j}\}=\{\textbf{r}_{j}^\text{e}\}$), \textit{ii}) the adiabatic approximation to be valid, permitting us to separate the high frequency vibrational modes of the molecule from its low frequency external translational and orientational modes, 
and \textit{iii}) the dynamical coupling between all of the molecular and the graphite degrees 
of freedom to be negligible.

The interaction potential energy $V_{\text{MG}}$ can then be written as
\begin{equation}
V_{\text{MG}}=V_{\text{MG}}(\textbf{r},\Omega ,\{Q_\nu ^{\text{e}}\})+V_{\text{%
MG}}(\{Q_\nu \},\textbf{r}^{\text{e}},\Omega ^{\text{e}}), \label{eq11}
\end{equation}
\noindent where $V_{\text{MG}}(\textbf{r},\Omega ,\{Q_\nu ^{\text{e}}\})$ represents the external motion-dependent part of the potential energy experienced by the non-vibrating molecule. 
$V_{\text{MG}}(\{Q_\nu \},\textbf{r}^{\text{e}},\Omega ^{\text{e}})$ characterizes the
vibrational dependent part, for the molecule at its equilibrium position and orientation;
in a perturbative approach it contributes to the molecular gas phase vibrational
Hamiltonian and leads to small vibrational frequency shifts.
\section{Adsorption observables}
At sufficiently low temperatures and within the rigid molecule and substrate
approximation, the potential energy surface $V_{\text{MG}}(X,Y)$, which is experienced by
the adsorbed molecule as it moves laterally above the surface can be calculated
by minimizing $V_{\text{MG}}$ with respect to both the perpendicular distance \textit{Z} between the
molecular centre of mass and the surface and the molecular angular coordinates ($\varphi$,$\theta$,$\chi$).

The resulting energy map gives information about the equilibrium site 
$(\mathbf{r}^{\textrm{e}},\Omega ^{\textrm{e}})$, and the lateral diffusion valley leading 
from one equilibrium site to an adjacent one. This makes it possible to determine the 
isosteric heat of adsorption or adsorption energy and also the surface diffusion constant.
\subsection{Adsorption energy}
The isosteric heat of adsorption $E_{\textrm{a}}$ of a single adsorbed molecule on the graphite
substrate can be defined as the energy required to keep the molecule with thermal energy 
$(\frac{n}{2}+1)kT$ at its equilibrium configuration on the surface of the substrate. It is
approximately expressed as
\begin{equation}
E_{\textrm{a}}\simeq \left( \frac n2+1\right) kT-V_{\text{MG}}^{\textrm{m}}(r^{%
\textrm{e}},\Omega ^{\text{e}})-\sum\limits_{\textrm{s=1}}^n\frac{\hbar \omega _{%
\textrm{s}}}2\coth \left( \frac{\hbar \omega _{\textrm{s}}}{2kT}\right), \label{eq12}
\end{equation}
\noindent where $n$ is the number of external degrees of freedom for the molecule (5 for linear
molecules and 6 for non linear molecules) and $k$ is the Boltzmann constant. The $\omega_{\text{s}}$ 
(s=1 to $n$) are the frequencies associated with the translational and orientational motions of
the molecule around their equilibrium positions.
\subsection{Surface diffusion constant}
Let $V_{\text{MG}}(\tau,\xi_{\gamma}(\tau))$ be the instantaneous potential energy surface 
experienced by the adsorbed molecule with mass $m$ as it moves along the diffusion valley described by the
coordinate path $\tau$. $\xi_{\gamma}(\tau)$ describe the instantaneous distance from the surface 
($\gamma = Z$) and the instantaneous orientation ($\gamma = \varphi,\theta,\chi$) of the molecule for the given $\tau$ value.

If we disregard the dynamics of the $\xi$ degrees of freedom around their $\tau$-dependent
equilibrium values $\hat{\xi}$ along the diffusion motion of the molecule and use the classical
transition state theory \cite{afvjdd1984}, then the jumping rate for the diffusion from one equilibrium
site S$_1$ to an equivalent adjacent one S$_2$ can be written as \cite{alcg1991}
\begin{equation}
K(T)\simeq \left( \frac{kT}{2\pi m^{*}}\right) ^{\frac 12}\frac{\exp \left[
-\Delta V_{\text{MG}}(\tau _{\text{0}})/kT\right] }{\int\limits_{\tau
_1}^{\tau _2}d\tau \exp \left[ -\Delta V_{\text{MG}}(\tau )/kT\right] }, \label{eq13}
\end{equation}
\noindent where $\Delta V_{\text{MG}}(\tau )=V_{\text{MG}}(\tau )-V_{\text{MG}}(\tau _1)$, and 
$\tau _{\text{0}}$ corresponds to the saddle point position along the diffusion valley. The diffusion constant is then
\begin{equation}
D(T)=\lambda ^{\text{2}}K(T), \label{eq14}
\end{equation}
\noindent where $\lambda =\tau _2-\tau _1$ is the jump length, i.e., the distance (in the diffusion direction) between the two adjacent equilibrium sites.

In Eq. (\ref{eq13}) $m^{*}$ is an effective molecular mass which is due to an inertial effect
resulting from the changes in the distance of the molecule from the surface and in the
molecule’s orientation as the diffusion process proceeds. It is expressed as
\begin{equation}
m^{*}=m\left( 1+\sum\limits_\gamma \frac{A_\gamma }m\hat{a}_\gamma ^2\right), \label{eq15}
\end{equation}
\noindent where $A_\gamma$ represents the molecular mass $m$ for $\gamma = Z$ and the molecular 
moments of inertia $I_a$, $I_b$, and $I_c$ for $\gamma$ = $\varphi$, $\theta$, and $\chi$, respectively.
The $\hat{a}_\gamma$ = $\frac{\partial\hat{\xi}_\gamma}{\partial\tau}$ represent deformation 
parameters as the molecule migrates from an equilibrium site to the saddle point position 
along the diffusion valley.
\section{Numerical results}
\subsection{Potential energy surfaces and equilibrium configurations}
The potential energy surfaces $V_{\text{MG}}(X,Y)$ for the CO, HCN, H$_2$O, H$_2$CO single molecules
adsorbed on the graphite substrate are presented in Figures 2 using a square surface
area of ($a \times \frac{2a}{\sqrt{3}}$), where the lattice parameter $a$ is shown in Figure 1.

In Table 3 we give, for each adsorbed species, some characteristics of the adsorption,
such as the most favourable positions and orientations and their associated energy
minima  $V_{\text{MG}}^{\text{m}}$, as well as the energy maxima $V_{\text{MG}}^{\text{M}}$.
\subsubsection{CO admolecule}
The potential energy surface experienced by the CO adsorbed molecule is presented 
in Figure 2a. The molecule exhibits an equilibrium configuration with an associated 
energy minimum $V_{\text{MG}}^{\text{m}}$ = -120.2 meV located above the hexagon site
centre at a distance $Z$ = 3.08 $\AA$ ̊from the surface and for a nearly flat orientation 
($\theta^{\text{e}}$ $\simeq$ 95 deg) position.

The perpendicular orientational motion around this equilibrium configuration is a
librational motion with an energy barrier height of about 40 meV, while that in the
plane parallel to the surface ($\varphi$) is a free rotational motion.

The energy maximum $V_{\text{MG}}^{\text{M}}$ = -106.5 meV is obtained with the molecule in a nearly
flat configuration at a distance $Z$ = 3.24 ̊$\AA$ above the graphite carbon sites. There is a
surface corrugation energy $V_{\text{MG}}^{\text{M}}-V_{\text{MG}}^{\text{m}}$ = 13.7 meV.

The saddle point of the diffusion path is reached when the molecular centre of
mass is above the middle of the carbon-carbon bonds at $Z$ = 3.20 ̊$\AA$ from the surface,
with the molecule perpendicular to the C-C bonds and being always in the nearly flat
configuration (Figure 3a). The associated energy barrier height is 
$\Delta V_{\text{MG}}(\tau_0)$ = 12 meV (Table 6).
\subsubsection{HCN admolecule}
The linear HCN adsorbed molecule presents a potential energy
surface with six most favourable sites per graphite hexagon (Figure 2b). The energy
minimum $V_{\text{MG}}^{\text{m}}$ = -175.5 meV is reached with the molecule in a nearly 
flat configuration ($\theta^{\text{e}}$ $\simeq$ 100 deg) with its centre of mass 
at $Z$ = 3.20 ̊̊$\AA$ on the half of the C-C bonds, with a displacement of 0.28 ̊$\AA$̊ 
inside the hexagon surface and with the hydrogen atom of the molecule pointing in 
the hexagon site centre direction.

In these equilibrium sites there is a strongly hindered orientational motion
perpendicular to the surface ($\theta$ motion) with an energy barrier height of about 75 meV
and a moderately hindered parallel one ($\varphi$ motion) with an energy barrier height of
about 16 meV.

However, it is interesting to note that this latter orientational motion, combined
with the translation motion of the molecule between the six equilibrium sites along a
quasi-circular trajectory of radius $\simeq$ 0.95 ̊$\AA$̊ around the hexagon site 
centre, could be regarded as a quasi-free motion (Figure 3b) with an energy 
barrier height $\leq$ 1.5 meV. Thus, an associated effective moment of inertia must 
be defined with respect to a new HCN y-axis lying at 0.95 ̊$\AA$̊ from its centre of 
mass (between the C and H atoms).

The energy maximum of $V_{\text{MG}}^{\text{M}}$ = -162.5 meV is obtained with the 
molecular centre of mass at $Z$ = 3.24 $\AA$̊, nearly above the hexagon site centre and 
also with the nearly flat configuration. This induces a surface corrugation energy of 13 meV.

Moreover, this position also corresponds to the saddle point of the diffusion motion
from one equilibrium site of a hexagon to one equivalent equilibrium site of an adjacent
hexagon (Figure 3b).

\begin{table} 
\begin{center}
\caption{Adsorption characteristics for CO, HCN, H$_2$O and H$_2$CO molecules.} 
\begin{tabular}{||c|c|c|c|c|c|c|c|c||}
\hline \hline
Molecule & $X^{\text{e}}$ (\AA ) & $Y^{\text{e}}$ (\AA ) & $Z^{\text{e}}$
(\AA ) & $\varphi ^{\text{e}}$ (deg) & $\theta ^{\text{e}}$ (deg) & $\chi ^{%
\text{e}}$ (deg) & $V_{\text{MG}}^{\text{m}}$ (meV) & $V_{\text{MG}}^{\text{M%
}}$ (meV) \\ 
\hline
CO & 0. & 0. & 3.08 & free & 95 & - & - 120.0 & - 106.5 \\ 
HCN & - 0.95 & 0. & 3.20 & 0 & 103 & - & - 175.5 & - 162.5 \\ 
H$_2$O & - 1.23 & - 0.36 & 3.00 & 90 & 100 & 90 & - 138.0 & - 129.5 \\ 
H$_2$CO & - 1.11 & 0. & 3.04 & 0 & 90 & 90 & - 208.5 & - 186.5 \\
\hline
\end{tabular}
\end{center}
\end{table}

\subsubsection{H$_2$O admolecule}
The potential energy surface of the non-linear H$_2$O molecule adsorbed on the graphite 
substrate is presented in figure 2c. The most favourable sites are obtained for the molecular 
centre of mass at $Z$ = 3.00 ̊$\AA$̊ above the C-C bonds, with a displacement of 0.36 $\AA$̊̊ from 
the C graphite carbon atom, the oxygen atom of the molecule being always close to this C graphite 
carbon atom (Figure 3c).

The energy minimum $V_{\text{MG}}^{\text{m}}$ = -138.0 meV is obtained with the molecule in a nearly
flat configuration ($\theta^{\text{e}}$ $\simeq$ 100 deg and $\chi$ = $\frac{\pi}{2}$) and sitting 
astride the C-C bonds (Table 3), and is in agreement with the available experimental 
value -156 meV obtained by Avgul and Kiselev \cite{nnaavk1970}.

It should be noted that in the equilibrium sites the angular motions of the C2 molecular 
symmetry axis (z-axis) parallel ($\varphi$ motion) and perpendicular ($\theta$ motion)
to the surface around their equilibrium values are moderately hindered motions with
energy barrier heights of about 8 and 15 meV, respectively. The angular (spinning)
motion about this axis ($\chi$ motion) is a completely hindered motion, the energy barrier
height being about 400 meV.

The energy maximum $V_{\text{MG}}^{\text{M}}$ = -129.5 meV is reached for $Z$ = 2.92 ̊$\AA$̊ above the 
hexagon site centre with the C$_2$ molecular symmetry axis perpendicular to the surface
of the graphite substrate. The surface corrugation energy is then 8.5 meV.

In Figure 3c we sketch the valley of diffusion of the water molecule on the surface.
The barrier height at the saddle point is $\Delta V_{\text{MG}}(\tau_0)$ = 1 meV (Table 6). 
Note however, that to obtain such a motion it is necessary to combine the translational motion and a
reorientational motion parallel to the surface ($\varphi$ = $\pm \frac{\pi}{3}$).

\begin{figure}
\includegraphics[width=18cm]{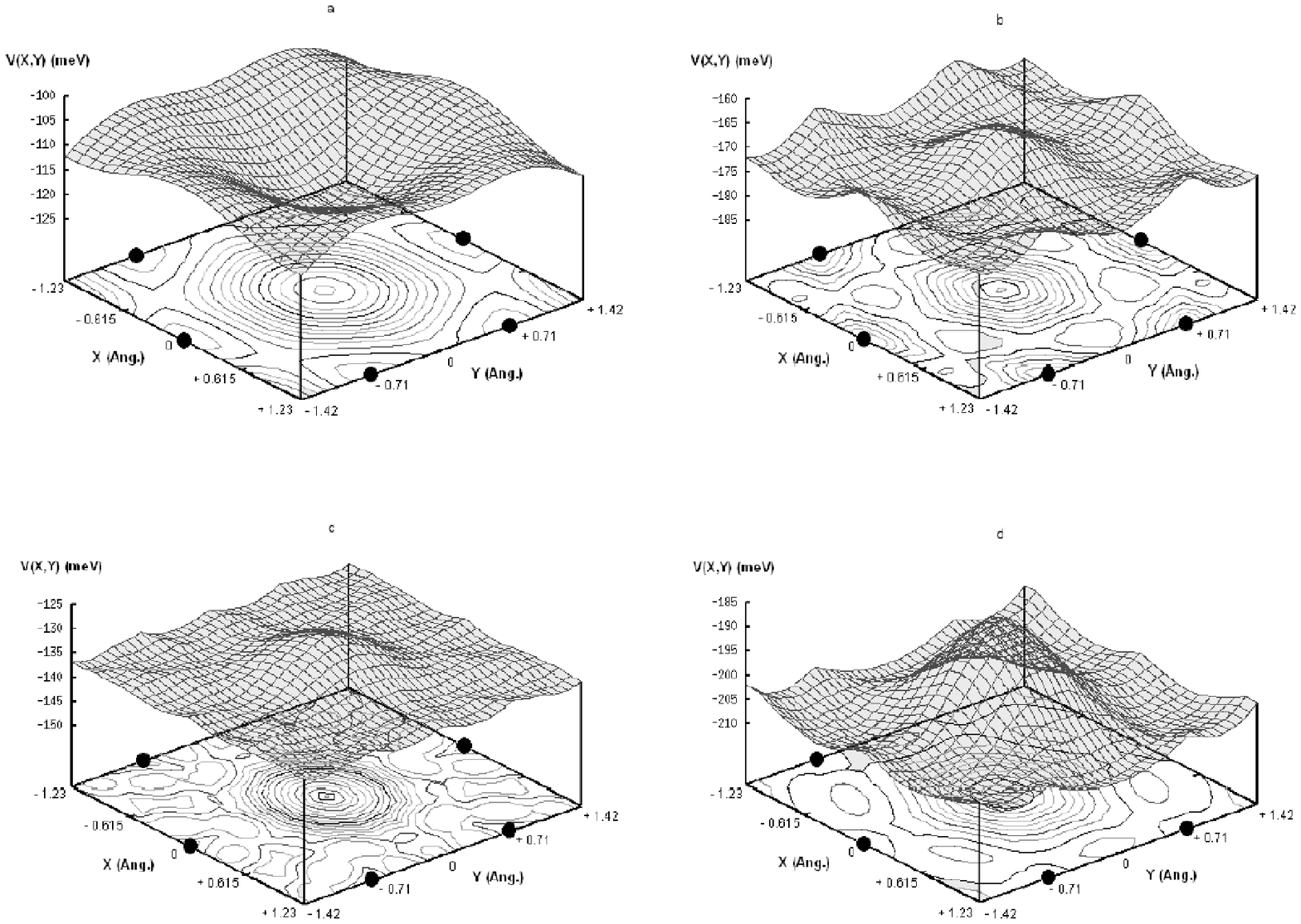}
\caption{The calculated potential energy surface for the adsorbed molecule on the graphite 
substrate. a (CO), b (HCN), c (H$_2$O) and d (H$_2$CO). Solid circles represent the carbon atoms 
of the substrate}
\end{figure}

\subsubsection{H$_2$CO admolecule}
Like the HCN molecule, the H$_2$CO molecule adsorbed on the graphite substrate exhibits six 
equilibrium sites per hexagon (Figure 2d). The energy minimum $V_{\text{MG}}^{\text{m}}$ = -208.5 meV 
is obtained for the molecule in a flat configuration ($\theta^{\text{e}}$ = 90 deg and 
$\chi^{\text{e}}$ = 90 deg) with its centre of mass at $Z$ = 3.04 ̊$\AA$ on half of the C-C bonds, 
with a displacement of 0.12 $\AA$ inside of the hexagon surface and with its C$_2$ symmetry axis (z-axis) pointing towards the hexagon site centre.

It should be noted that in its equilibrium sites H$_2$CO behaves nearly like HCN with
regard to the perpendicular $\theta$ and parallel $\varphi$ angular motions of its z-axis, in one hand,
but like H$_2$O with regard to its spinning motion about this axis.

The energy maximum is $V_{\text{MG}}^{\text{M}}$ = -186.5 meV, with the molecule in the flat
configuration at $Z$ = 3.12 $\AA$̊ above the hexagon site centre. The surface corrugation 
energy is then 22 meV.

The saddle point of the diffusion path is obtained for the molecule in the flat
configuration at $Z$ = 3.04 $\AA$̊ above the C carbon site with a corresponding energy
barrier height $\Delta V_{\text{MG}}(\tau_0)$ = 5 meV (Table 6). Moreover, the rotation of 
the molecule parallel to the surface plane is a free motion on this site.

We note that the fact that the equilibrium configuration is nearly above the carbon-
carbon bonds for HCN, H$_2$O, H$_2$CO molecules results from their strong multipole
moments. Their quantum $V_{\text{LJ}}$, induction $V_{\text{I}}$ and electrostatic $V_{\text{E}}$ 
contribution parts represent $\sim$ 80$\%$, $\sim$ 15$\%$ and $\sim$ 5$\%$, respectively, of 
the total potential energy for these molecules.

The weakly polar CO molecule behaves as a quadrupolar molecule and adopts an
equilibrium configuration above the hexagon site centre. Its quantum contribution part
$V_{\text{LJ}}$ represents more than $\sim$ 95$\%$ of the total potential energy.
\subsection{Adsorption energy calculations}
The calculation, using Eq. (\ref{eq12}), of the adsorption energy values for single molecules
adsorbed on the graphite substrate requires a knowledge of the frequencies $\omega _{\text{s}}$
associated with the linear and angular oscillations associated with the translational
and orientational motions of the molecules around their equilibrium configurations on
the most favourable sites.

A discrete variable representation method \cite{jcliphjvl1985} was used to solve the Schr̈\"{o}dinger 
equation giving the energy levels of a molecule undergoing linear and angular oscillations
at its equilibrium site, with the other degrees of freedom being assumed to remain at
their equilibrium values. The various frequencies calculated for the molecules treated in
this work are reported in table 4.

\begin{table} 
\caption{Calculated frequencies (cm$^{\text{-1}}$) associated with the translational and orientational 
oscillations of CO, HCN, H$_2$O and H$_2$CO molecules.} 
\begin{center}
\begin{tabular}{||c|c|c|c|c|c|c||}
\hline \hline
Molecule & $\omega _{\text{x}}$ & $\omega _{\text{y}}$ & $\omega _{\text{z}}$
& $\omega _\varphi $ & $\omega _\theta $ & $\omega _\chi $ \\ 
\hline
CO & 26.6 & 26.6 & 62.1 & $\sim $ 0 & 76.6 & - \\ 
HCN & 21.0 & 23.4 & 79.0 & 32.3 & 109.7 & - \\ 
H$_2$O & $\sim $ 0 & $\sim $ 0 & 90.3 & 22.6 & 35.5 & 233.9 \\ 
H$_2$CO & 25.0 & $\sim $ 0 & 155.7 & 21.0 & 91.1 & 203.3 \\
\hline
\end{tabular}
\end{center}
\end{table}

As can be expected from the potential energy surfaces, the frequencies associated
with the orientational motions perpendicular to the surface ($\theta$ and $\chi$ motions) 
(except the $\theta$ motion for H$_2$O molecule) are large compared to those associated with the
orientational motion parallel to the surface ($\varphi$ motion).

Finally, the calculated frequencies were introduced into Eq. (\ref{eq12}) and the adsorption
energies were determined for three typical temperatures. The resulting values are
reported in table 5.

\begin{table} 
\begin{center}
\caption{Calculated adsorption energy $E_{\text{a}}$ (meV) of CO, HCN, H$_2$O and H$_2$CO 
molecules for three typical temperatures.} 
\begin{tabular}{||c|c|c|c||}
\hline \hline
Molecule & $T$ = 10 K & $T$ = 40 K & $T$ = 100 K \\ 
\hline
CO & 111.3 & 114.8 & 114.4 \\ 
HCN & 168.5 & 163.8 & 159.5 \\ 
H$_2$O & 117.6 & 123.7 & 129.6 \\ 
H$_2$CO & 181.0 & 186.6 & 189.8 \\
\hline
\end{tabular}
\end{center}
\end{table}

\subsection{Diffusion constant calculations}
The study of the potential energy surface for each adsorbed species allows us to
determine the diffusion path, its length $\lambda$ and potential energy function 
$\Delta V_{\text{MG}}(\tau)$, and the deformation parameters $\hat{a}_\gamma$ as the molecular 
diffusion proceeds on the surface.

It must be mentioned that large values for $\hat{a}_\gamma$ correspond to important equilibrium
changes when the molecular centre of mass migrates from the minimum to the saddle
point position along the diffusion valley. For instance, the angular deformation
parameter values $\hat{a}_\varphi$ are 1.03 and 0.74 \textit{rad}/$\AA$̊ for H$_2$O and H$_2$CO admolecules,
respectively.

In contrast, the angular deformation parameters $\hat{a}_\theta$ and $\hat{a}_\chi$ are negligibly 
small for all the admolecules because of the very weak changes of their perpendicular 
orientation with respect to the surface during the diffusion motion.

The linear deformation parameter values $\hat{a}_z$ are 0.10, 0.05, 0.04 and 0.00 for 
CO, HCN, H$_2$O and H$_2$CO admolecules, respectively.

Finally, the possible diffusion paths are indicated in figures 3 and the corresponding
energy barrier heights $\Delta V_{\text{MG}}(\tau_0)$ are given in table 6.

\begin{figure}
\includegraphics[width=15cm]{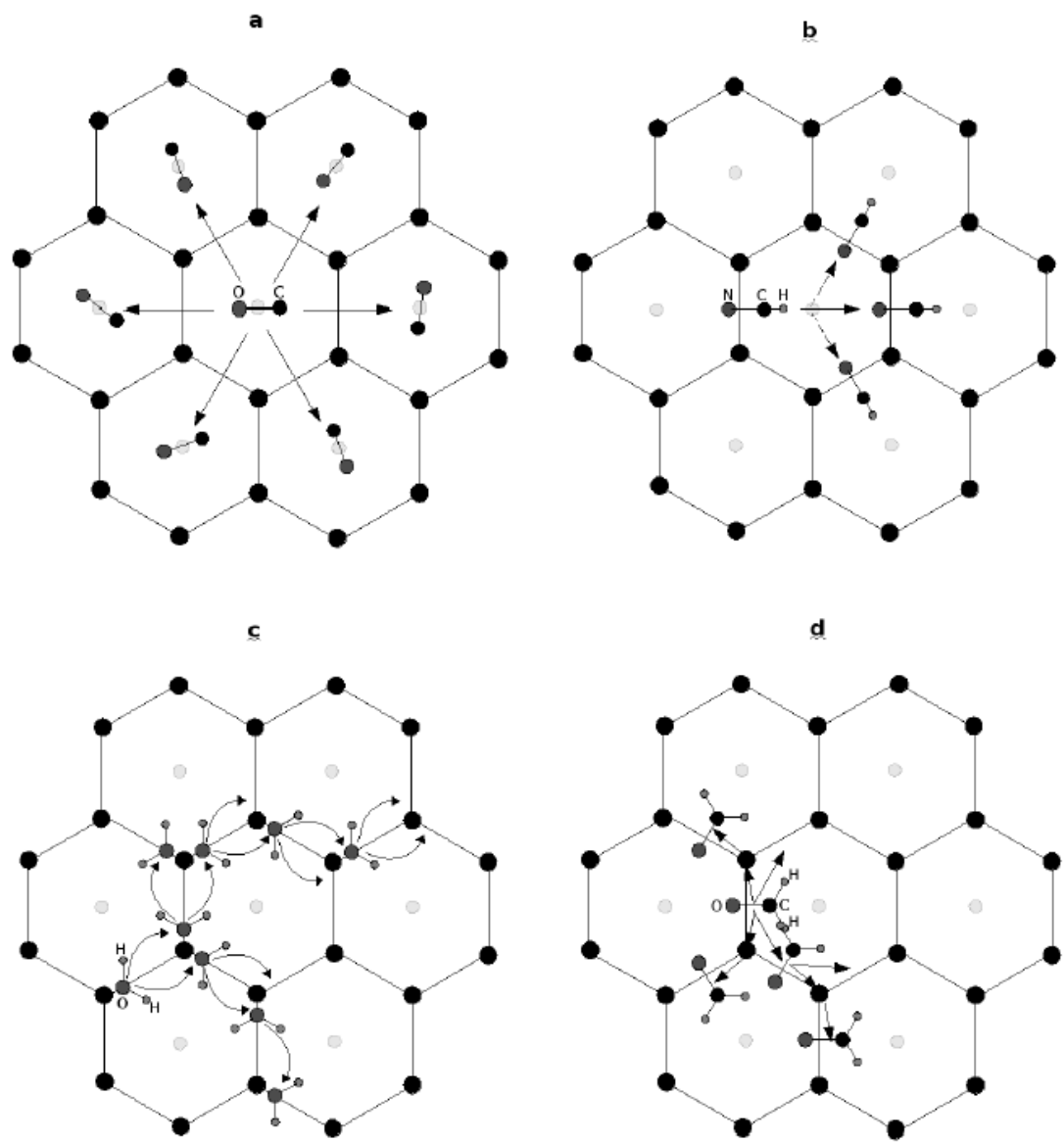}
\caption{The calculated equilibrium positions for the adsorbed molecule on the graphite 
substrate, indicating the energetically favoured paths for diffusion between the equilibrium 
positions. a (CO), b (HCN), c (H$_2$O) and d (H$_2$CO).}
\end{figure}

Introducing the calculated deformation parameters in Eq. (\ref{eq15}) the mass ratios $\dfrac{m^*}{m}$ 
are obtained and presented in table 6. Note that for the CO and HCN admolecules the mass changes are negligibly small, while for H$_2$O and H$_2$CO the effective mass increases by 11$\%$ and 
27$\%$, respectively.

\begin{table} 
\begin{center}
\caption{Calculated surface diffusion constant $D$ (cm$^{\text{2}}$/s) of CO, HCN, H$_2$O and H$_2$CO 
molecules for three typical temperatures.} 
\begin{tabular}{||c|c|c|c|c|c||}
\hline \hline
Molecule & $\frac{m^{*}}m$ & $\Delta V_{\text{MG}}(\tau _0)$ (meV) & $T$ =
10 K & $T$ = 40 K & $T$ = 100 K \\ 
\hline
CO & 1.01 & 12. & 3.3 $\times 10^{-10}$ & 1.  $\times 10^{-5}$ & 7.3  $%
\times 10^{-5}$ \\ 
HCN & 1.002 & 13. & 1.  $\times 10^{-10}$ & 1.2 $\times 10^{-5}$ & 7.2 $%
\times 10^{-5}$ \\ 
H$_2$O & 1.11 & 1. & 1.5 $\times 10^{-5}$ & 4.3 $\times 10^{-5}$ & 7.2 $%
\times 10^{-5}$ \\ 
H$_2$CO & 1.27 & 5. & 2.2 $\times 10^{-7}$ & 1.8 $\times 10^{-5}$ & 5.1 $%
\times 10^{-5}$ \\ 
\hline
\end{tabular}
\end{center}
\end{table}

\section{Discussion}
In the present work we have studied the adsorption of several molecules on a graphite
substrate; the molecules were selected because of their biophysical interest, as explained
in the introduction.

The major calculation was that of the potential energy surface for each molecule and
the figures show the resulting surfaces. A minimization calculation then gave the possible
equilibrium positions for each molecule, while the potential energy surface indicated the
favourable pathways between these positions which would facilitate surface diffusion of
the adsorbed molecule.

The calculated adsorption energies as shown in table 5 show only a relatively
weak variation with temperature, although the particular values of the molecule-surface
interaction parameters for the HCN molecule lead to a temperature dependence which
is in the opposite direction to that for the other three molecules studied.

The calculated diffusion constants for the molecules are shown in table 6 and it
is clear that the different saddle point barrier heights and effective masses lead to a
variation of several orders of magnitude between the low temperature diffusion constants
of the various molecules. In effect, one can remark that at \textit{T} = 10 K the H$_2$O 
molecule could diffuse 3 orders of magnitude more easily than H$_2$CO and 5 orders of magnitude 
more easily than CO and HCN molecules.

However, as the temperature increases the thermal energy \textit{kT} becomes sufficiently 
large to overcome the barriers for all the molecules, with the consequence that the diffusion 
constants vary much less from molecule to molecule as the temperature rises.

\section*{ACKNOWLEDGEMENTS}

The authors gratefully acknowledge fruitful discussions with Drs. D. Viennot and S. Picaud.

\section*{Appendix : Rotational matrix transformation}
%\appendix{Rotational matrix transformation}
%
The unitary matrix \textbf{M} characterizing the transformation from the surface absolute frame 
(\textbf{O,X,Y,Z}) into the molecular one (\textbf{G,x,y,z}) through the Euler angles $\varphi$, 
$\theta$ and $\chi$ is given by \cite{mer1967}

\begin{eqnarray}
\textbf{M}(\varphi ,\theta ,\chi ) = \left( 
\begin{array}{ccc}
\cos \varphi \text{ }\cos \theta \text{ }\cos \chi \text{ - }\sin \varphi 
\text{ }\sin \chi  & \sin \varphi \text{ }\cos \theta \text{ }\cos \chi 
\text{ + }\cos \varphi \text{ }\sin \chi  & \text{- }\sin \theta \text{ }%
\cos \chi  \\ 
\text{- }\cos \varphi \text{ }\cos \theta \text{ }\sin \chi \text{ - }\sin
\varphi \text{ }\cos \chi  & \text{- }\sin \varphi \text{ }\cos \theta \text{
}\sin \chi \text{ + }\cos \varphi \text{ }\cos \chi  & \sin \theta \text{ }%
\sin \chi  \\ 
\cos \varphi \text{ }\sin \theta  & \sin \varphi \text{  }\sin \theta  & 
\cos \theta \text{ }
\end{array}
\right)  \nonumber
\end{eqnarray}

\end{document}